\documentclass[prl,twocolumn,showpacs,groupedaddress]{revtex4}

\usepackage{graphicx}

\begin{document}

\title{Dephasing of Atomic Tunneling by Nuclear Quadrupoles}
\author{Alois W\"{u}rger$^{1}$, Andreas Fleischmann$^{2}$, Christian Enss$^{2}$}

\address{$^{1}$Universit\'{e} Bordeaux 1, CPMOH\thanks{Unit\'{e} Mixte de Recherche CNRS 5798}, 351 cours de la
Lib\'{e}ration, 33405 Talence, France\\
$^{2}$Kirchhoff-Institut f\"{u}r Physik, Universit\"{a}t Heidelberg,
INF 227, 69120 Heidelberg, Germany}

\date{\today }

\begin{abstract}
Recent experiments revealed a most surprising magnetic-field dependence of
coherent echoes in amorphous solids. We show that a novel dephasing
mechanism involving nuclear quadrupole moments is the origin of the observed
magnetic-field dependence.
\end{abstract}


\maketitle

Until recently it was the general believe that the dielectric
properties of insulating glasses -- free of magnetic impurities --
are largely independent of external magnetic fields. New
investigations, however, have shown that the dielectric properties
of certain multi-component glasses at very low temperatures are
strongly influence by a magnetic field \cite
{Str98,Str00,Woh01,Hau02,Coc02,Lud02,Ens02}. In particular, the
low-frequency dielectric susceptibility and the amplitude of
spontaneous polarization echoes generated in these amorphous
materials show a striking non-monotonic dependence on applied
magnetic field.

Since the low-temperature properties of glasses are governed by atomic
tunneling systems (for recent reviews see \cite{Esq98,Hun00}), it has been
speculated whether and how a magnetic field can couple to quantum tunneling.
Two models have been proposed, that relate the magnetic-field dependence to
the Ahanorov-Bohm phase of a charged particle moving along a closed loop
\cite{Ket99,Wue02}. Very recent polarization echo experiments, however,
indicate that such a periodic variation of the tunnel splitting is not the
origin of the observed magnetic field effects \cite{Lud02,Ens02}. In
contrast, these experiments strongly suggest that nuclear magnetic moments
play a crucial in the observed anomalies.

In this paper we discuss how nuclear magnetic and quadrupolar moments
influence atomic tunnel states. After a brief reminder of two-pulse echoes
of two-level systems and the nuclear spin hamiltonian, we give the
echo-amplitude correction factor due to nuclear spins and evaluate this
expression for the limiting cases of weak and strong magnetic fields.
Finally, we compare our theory with recent data for several glasses.

\begin{figure}[t]
\includegraphics[width=0.8\linewidth]{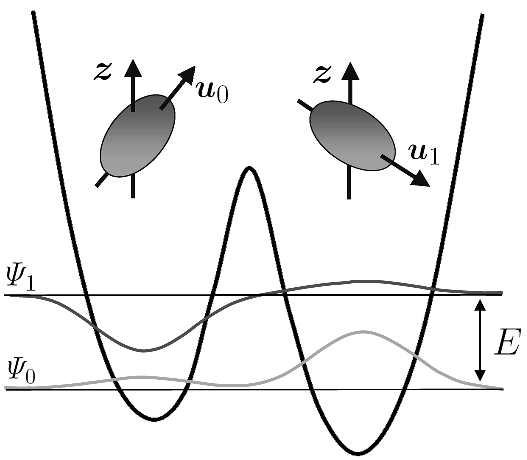} \vskip-2mm
\caption{Two-level system with energy splitting $E$, eigenfunctions $\protect%
\psi _{0}$ and $\protect\psi _{1}$, and the corresponding quadrupole
quantization axes $\mathbf{u}_{0}$ and $\mathbf{u}_{1}$.}
\label{figure1}
\end{figure}

Two-level systems (TLS) in glasses arise from double-well potentials with
asymmetry $\Delta $ and tunnel matrix element $\Delta _{0}$,
\begin{equation}
H=\mathbf{-}\frac{1}{2}\Delta _{0}\sigma _{x}-\frac{1}{2}\Delta \sigma _{z},
\label{eq2}
\end{equation}
where $\sigma _{z}=\pm 1$ is the reduced two-state variable that indicates
the states localized in the two wells \cite{Esq98,Hun00}. The eigenstates $%
\psi _{0}$ and $\psi _{1}$\ are separated by the energy splitting $E=\sqrt{%
\Delta _{0}^{2}+\Delta ^{2}}$ (see Fig.\ 1.) In a two-pulse echo experiment,
the first pulse creates a coherent superposition of the ground state and
excited state, with a relative phase factor $e^{-iEt/\hbar }$.\ Because of
the dispersion of the two-level splitting $E$, the corresponding macroscopic
polarization decays rapidly. After a waiting time $t_{\mathrm{w}}$, the
second pulse exchanges the amplitudes of these two states; the resulting
phase factor $e^{-iE(t-t_{\mathrm{w}})/\hbar }$ leads to a revival of the
coherent polarization, and the \ ``echo'' is observed at a time $t\approx t_{%
\mathrm{w}}$ after the second pulse,
\begin{equation}
P_{0}(t,t_{\mathrm{w}})=\sum_{i}A_{i}\cos \left[ \omega _{i}(t-t_{\mathrm{w}%
})\right] ,  \label{eq4}
\end{equation}
where the sum runs over all TLS with tunnel frequency $\omega
_{i}=E_{i}/\hbar $\ and effective dipole moment $A_{i}$.

Such a tunnel system involves several atoms, each of which may carry a
nuclear magnetic dipole and an electric quadrupole.\ For the sake of
simplicity, we consider a single atom whose nucleus is in a state of total
angular momentum $\mathbf{I}$ where $\mathbf{I}^{2}=\hbar ^{2}I(I+1)$. This
``nuclear spin'' results in a magnetic moment $g\mu _{\mathrm{N}}\mathbf{I}%
/\hbar $, with the Land\'{e} factor $g$ and the nuclear magneton $\mu _{%
\mathrm{N}}=5\times 10^{-27}$ J/T.\ In the case $I\geq 1$, the orbital
motion of the protons is related to an electric quadrupole moment \cite
{Bla58}; for a nuclear charge distribution $\rho (\mathbf{r})$\ oriented
along the axis $\mathbf{e}$ one finds
\begin{equation}
Q=\int d^{3}r\left[ 3(\mathbf{r\cdot e})^{2}-\mathbf{r}^{2}\right] \rho (%
\mathbf{r}).  \label{eq6}
\end{equation}
The magnetic dipole couples to the external field $\mathbf{B}=B\mathbf{e}_{z}
$ and the quadrupole moment to the electric field gradient (EFG) that is
given by the curvature of the crystal field potential $\phi (\mathbf{r})$.
We consider the simplest case of a single diagonal term$\ \phi ^{\prime
\prime }=(\mathbf{u\cdot \nabla })^{2}\phi $ along the axis $\mathbf{u}$.
Then the spin hamiltonian reads as \cite{Bla58}
\begin{equation}
V=g\mu _{\mathrm{N}}B\hat{I}_{z}+\frac{\phi ^{\prime \prime }Q}{4}\frac{3%
\hat{I}_{u}^{2}-I(I+1)}{I(2I-1)},  \label{eq8}
\end{equation}
with the projections of the nuclear spin operator on the axes defined by the
EFG, $\hat{I}_{u}=(\mathbf{u\cdot I})/\hbar $, and the magnetic field, $\hat{%
I}_{z}=I_{z}/\hbar $.

For zero magnetic field, $\mathbf{u}$ is the appropriate quantization axis;
with $\hat{I}_{u}^{2}=m^{2}$ and $m=-I,...,I$ the hamiltonian is diagonal. ($%
I=1$ and $I=3/2$ give rise to a doublet, $I=2$ and $I=5/2$ to a triplet,
etc.) In the opposite case of zero EFG, the usual choice $\hat{I}_{z}=m$
gives the $(2I+1)$ Zeeman levels $mg\mu _{\mathrm{N}}B$. In general, the
axes defined by the magnetic field, $\mathbf{e}_{z}$, and the EFG, $\mathbf{u%
}$, are not parallel, i.e., the operators $\hat{I}_{u}$ and $\hat{I}_{z}$
cannot be diagonalized simultaneously, thus resulting in a more complicated
situation if both the magnetic field and the EFG are finite.

For asymmetric TLS, one of the eigenstates, say the ground state $\psi _{0}$%
, has a large probability amplitude in the left well, whereas the excited
one, $\psi _{1}$, has a larger amplitude in the right well (see Fig. 1.) In
an amorphous or disordered solid, the crystal field, and thus the quadrupole
quantization axis, are not the same in the two wells. In terms of the
nuclear spin hamiltonian, this means that both the absolute value of the EFG
and the quantization axis depend on the two-state variable. These quantities
are denoted $\phi _{0}^{\prime \prime }$ and $\mathbf{u}_{0}$ in the ground
state, and $\phi _{1}^{\prime \prime }$ and $\mathbf{u}_{1}$ in the excited
level.

Now we discuss how nuclear spins affect the polarization echo that arises
from a coherent superposition of the two tunnel states. In general, the
quadrupolar part of the nuclear spin hamiltonian does not commute with $H$.
Thus $V$ leads to a dispersion of the two-level splitting when switching
from $\ \mathbf{u}_{0}$ to $\mathbf{u}_{1}$ during the two pulses, and thus
to a dephasing of the the echo signal.

The density operator of a TLS involves four independent operators, e.g., the
three Pauli matrices $\sigma _{x}$, $\sigma _{y}$,\ $\sigma _{z}$, and
unity. Accordingly, the propagator is represented by a four-dimensional
matrix \cite{Wue93}. Taking into account a nuclear spin $I$ renders the
dynamics significantly more complex, since we have to deal with $D=2(2I+1)$
quantum states corresponding to a density matrix with $D^{2}$ entries. It
can be shown that a nuclear spin results in an overall factor of the echo
amplitude \cite{Wue-unp},

\begin{equation}
P(t,t_{\mathrm{w}})=P_{0}(t,t_{\mathrm{w}})f(t,t_{\mathrm{w}}),  \label{eq10}
\end{equation}
where $f(t,t_{\mathrm{w}})$ is determined by the nuclear spin energies and
eigenfunctions in the upper and lower tunnel states,
\begin{equation}
f(t,t_{\mathrm{w}})=\overline{\;\sigma _{z}e^{-i\mathcal{L}t}\mathcal{R}%
(\Omega \tau _{2})e^{-i\mathcal{L}t_{\mathrm{w}}/\hbar }\mathcal{R}(\Omega
\tau _{1})\;}.  \label{eq12}
\end{equation}
The bar indicates the ensemble average; time evolution for zero driving
field is written in terms of the Liouville operator $\mathcal{L}\ast
=(1/\hbar )[V,\ast ]$, and the rotations $\mathcal{R}$ account for the two
electric-field pulses of duration $\tau _{i}$ and Rabi frequency $\Omega $.
Both $\mathcal{L}$ and $\mathcal{R}$\ are superoperators that act on nuclear
spin variables. The argument of $\mathcal{R}(\theta )$ is the \ ``pulse
area'' $\theta $.

The first external-field pulse creates a coherent superposition of the two
tunnel states, whereas the second pulse exchanges their phases. (In the
Bloch spin picture, this is related to rotations of the ``polarization
vector'' $\left\langle \overrightarrow{\sigma }\right\rangle $.) The full
density matrix is expressed through standard basis operators $\left| \mathbf{%
u}_{i}\alpha \right\rangle \left\langle \mathbf{u}_{j}\beta \right| $,\ with
$i,j=0,1$\ and $\alpha ,\beta =-I,...,I$. In the composite space of TLS and
nuclear spin variables, $\mathcal{L}$ and $\mathcal{R}$ are represented by
tetrads $L_{pqrs}$ and $R_{pqrs}$ of dimension $D^{2}$. Spelling out the
matrices $\mathcal{R}$ and $\sigma _{z}$ and taking the trace, we obtain the
correction factor
\begin{equation}
f(t,t_{\mathrm{w}})=\sum_{\alpha \beta \gamma \delta }\overline{f_{\alpha
\beta \gamma \delta }\cos \left[ \varepsilon _{0\alpha }-\varepsilon
_{1\beta })t/\hbar -\left( \varepsilon _{0\gamma }-\varepsilon _{1\delta
}\right) t_{\mathrm{w}}/\hbar \right] },
\end{equation}
where nuclear spin energy levels (i.e. the eigenvalues of $V$) are denoted
by $\varepsilon _{0\alpha }$\ and $\varepsilon _{1\alpha }$ and $f_{\alpha
\beta \gamma \delta }$ depends on the matrix elements of  $\mathcal{R}$ and $%
\sigma _{z}$. (Details will be given elsewhere \cite{Wue-unp}.) Here we
resort to a simple approximation that is justified in various situations,
such as short pulses or almost parallel quadrupolar quantization axes, and
that is expected to grasp the essential physics in any case. Since the
polarization echo occurs on a time scale much shorter than the waiting time,
we may put $t=t_{\mathrm{w}}$.\ For a TLS initially in the ground state we
the phase factor,
\begin{equation}
f(t_{\mathrm{w}})=\frac{1}{2I+1}\sum_{\alpha ,\beta ,\delta }\overline{|%
\mathcal{\chi }_{\alpha \beta }|^{2}|\mathcal{\chi }_{\alpha \delta
}|^{2}\cos (\omega _{\beta \delta }t_{\mathrm{w}})},  \label{eq20}
\end{equation}
that depends on the overlaps
\begin{equation}
\mathcal{\chi }_{\alpha \beta }=\left\langle \mathbf{u}_{0}\alpha |\mathbf{u}%
_{1}\beta \right\rangle   \label{eq16}
\end{equation}
and the\ quadrupole spectrum,

\begin{equation}
\omega _{\beta \delta }=\left( \varepsilon _{1\delta }-\varepsilon _{1\beta
}\right) /\hbar .
\end{equation}
For the case where the system is initially in the excited state, we find a
similar expression with $\varepsilon _{0\gamma }$ instead of $\varepsilon
_{1\gamma }$. Thus $f(t_{\mathrm{w}})$ describes the reduction of the whole
echo signal.

In the remainder of this paper, we discuss the reduction factor (\ref{eq20}%
)\ First we consider the situation where the quadrupole coupling is
ineffective, such as a zero EFG, parallel quantization axes $\mathbf{u}_{0}=%
\mathbf{u}_{1}$, or a very strong magnetic field. Then the nuclear spin
states corresponding to the tunnel levels\ are identical, $\left| \mathbf{u}%
_{0}\alpha \right\rangle =\left| \mathbf{u}_{1}\alpha \right\rangle $, for $%
\alpha =-I,...,I$, and $\mathcal{\chi }_{\alpha \beta }=\delta _{\alpha
\beta }$, resulting in $f(t_{\mathrm{w}})\equiv 1$.

In general, however, the EFG is finite and the quadrupolar hamiltonian is
not the same for the two levels, resulting in a non-diagonal overlap matrix,
$\mathcal{\chi }_{\alpha \beta }$. The quadrupolar energy scale reads as
\begin{equation}
\hbar \omega _{\mathrm{Q}}=\frac{3}{4I(2I-1)}\phi ^{\prime \prime }Q.
\end{equation}
Typical values for the quadrupolar energy $\phi ^{\prime \prime }Q$
correspond to frequencies of the order of tens of MHz and thus satisfy, for
waiting times $t_{\mathrm{w}}\sim \mu \sec $, the inequality $\omega _{%
\mathrm{Q}}t_{\mathrm{w}}\gg 2\pi $ that simplifies significantly the
analysis. As a consequence, all terms involving different quadrupole levels $%
\gamma \neq \pm \beta $ in (\ref{eq20}) vanish. Yet note that the
quadrupolar spectrum exhibits a degeneracy with respect to $\beta
\rightarrow -\beta $ which, in turn, is lifted by a magnetic field.

In the limit of zero waiting time the cosines in (\ref{eq20}) are equal to
unity, and we have $f(t_{\mathrm{w}}\rightarrow 0)=1$.\ In the opposite case
of very long times and all degeneracies lifted, the terms with finite
frequency $\omega _{\gamma \beta }$ vanish, resulting in
\begin{equation}
f(t_{\mathrm{w}}\rightarrow \infty )=\frac{1}{2I+1}\sum_{\alpha ,\beta }%
\overline{|\mathcal{\chi }_{\alpha \beta }|^{4}}=1-a.
\end{equation}
We are interested in the intermediate regime of experimentally relevant
waiting times that are of the order of $\mu \sec $.\ Thus we have to look
for frequencies in the MHz range that satisfy the condition $\omega _{\gamma
\beta }t_{\mathrm{w}}\sim \pi $.

Both the overlaps $\mathcal{\chi }_{\alpha \beta }$ and the frequencies $%
\omega _{\gamma \beta }$\ depend in an intricate manner on the relevant
orientation of the three vectors $\mathbf{e}_{z}$, $\mathbf{u}_{0}$, $%
\mathbf{u}_{1}$ and on the ratio of the Zeeman splitting and the quadrupolar
energy. Here we discuss a few limiting cases where (\ref{eq20}) simplifies
significantly. The argument is developed for half-integer spin $I=\frac{3}{2}%
,\frac{5}{2},...$ but easily generalized to integer $I$.

First we consider the case of a weak magnetic field where
\begin{equation}
\hbar \omega _{\mathrm{Z}}=g\mu _{\mathrm{N}}B
\end{equation}
is small as compared to $\hbar \omega _{\mathrm{Q}}$. Then the nuclear
Zeeman splitting $\hbar \omega _{\mathrm{Z}}$ lifts the degeneracy of the
doublets $\pm \beta $ of the quadrupolar energy ($\beta =\frac{1}{2},...,I$%
). Discarding rapidly oscillating terms $\sim \cos \left( \omega _{\mathrm{Q}%
}t_{\mathrm{w}}\right) $, and separating the weight factor
\[
b_{\beta }=2\sum_{\alpha }\overline{|\mathcal{\chi }_{\alpha \beta }|^{2}|%
\mathcal{\chi }_{\alpha ,-\beta }|^{2}}
\]
and the time-dependent term, we obtain \
\begin{equation}
f(t_{\mathrm{w}})=1-a+\sum_{\beta }b_{\beta }\overline{\cos (\omega _{\beta
,-\beta }t_{\mathrm{w}})}.  \label{eq30}
\end{equation}
For small $B$ we may neglect the magnetic-field dependence of the weight
factors and treat the Zeeman term as a perturbation.\ Starting from the
eigenbasis of the quadrupolar energy, we thus diagonalize the Zeeman term of
$V$ in each degenerate subspace $\pm \beta $. The splitting of each doublet $%
\pm \beta $ depends on the relative orientation of the quantization axes $%
\mathbf{e}_{z}$ and $\mathbf{u}_{1}$ through the cosine $x=\mathbf{e}%
_{z}\cdot \mathbf{u}_{1}$. For $\ \beta =\frac{3}{2},...,I$, the splitting
is given by the projection of the magnetic field on the quadrupolar axis,\
\begin{equation}
\omega _{\beta ,-\beta }=x2\beta \omega _{\mathrm{Z}}\;\;\;\;(\beta >1/2),
\end{equation}
whereas for $\beta =\frac{1}{2}$ it reads as
\begin{equation}
\omega _{\frac{1}{2},-\frac{1}{2}}=\sqrt{x^{2}+\left( I+1/2\right)
^{2}\left( 1-x^{2}\right) }\omega _{\mathrm{Z}}.
\end{equation}
The average in (\ref{eq30}) is given by
\begin{equation}
\overline{\left( ...\right) }=\int_{0}^{1}dxp(x)\left( ...\right) .
\label{eq34}
\end{equation}

\begin{figure}[t]
\includegraphics[width=0.92\linewidth]{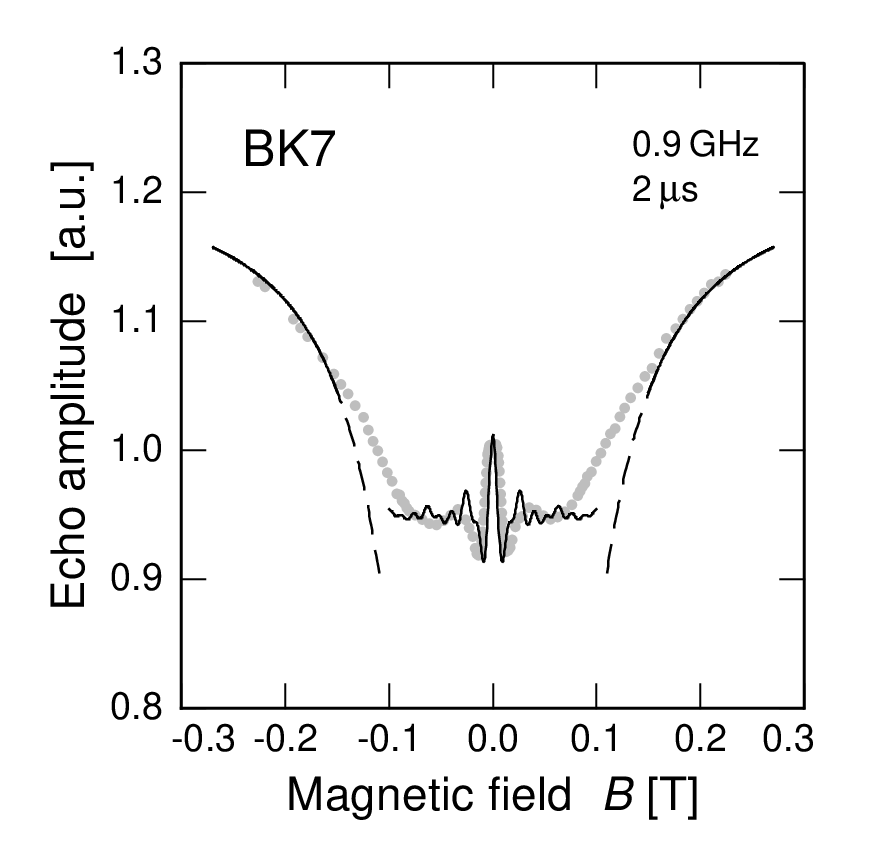} \vskip-5mm
\caption{Magnetic-field dependence of the echo amplitude of BK7 \protect\cite
{Lud-unp}. The solid lines are calculated as explained in the text. }
\end{figure}

Though possible in principle, calculation of the normalized distribution $%
p(x)$ is beyond the scope of the present paper. Eq. (\ref{eq30}) shows
oscillatory behavior with period $\sim I\omega _{\mathrm{Z}}t_{\mathrm{w}}$,
independent of the precise form of $p(x)$. For the fit of the experimental
data in Fig. 2, the best results are obtained with $p(x)=3x^{2}$, and $b_{%
\frac{1}{2}}=2b_{\frac{3}{2}}$. Yet note that the position of the first
minimum of $f(t_{\mathrm{w}})$ occurs always at $\omega _{\mathrm{Z}}t_{%
\mathrm{w}}\approx 0.6\pi $ and hardly depends on $p(x)$.

Now we turn to strong magnetic fields, $\omega _{\mathrm{Z}}\gg \omega _{%
\mathrm{Q}}$, where the quadrupole energy may be treated as a perturbation
with respect to the Zeeman splitting. When calculating the overlap matrix $%
\chi _{\alpha \beta }$ to lowest order in $\omega _{\mathrm{Q}}/\omega _{%
\mathrm{Z}}$ and observing the normalization condition $\sum_{\beta }|\chi
_{\alpha \beta }|^{2}=1$, we obtain \
\begin{equation}
1-f(t_{\mathrm{w}})=2\sum_{\alpha ,\beta \neq \alpha }|\chi _{\alpha \beta
}|^{2}\sim \left( \frac{\hbar \omega _{\mathrm{Q}}}{\mu _{\mathrm{N}}B}%
\right) ^{2}\; .  \label{eq40}
\end{equation}
Thus we find a variation $f(t_{\mathrm{w}})=1-\mathrm{const}.\times B^{-2}$
at high magnetic fields, as shown in Fig.\ 2.

These theoretical findings (\ref{eq30}--\ref{eq40}) agree rather well with
available data. We briefly discuss the most salient features.

(i) Both the oscillatory behavior of the echo amplitude with small $B$ and
the saturation at higher fields have been observed for several
multicomponent glasses and mixed crystals \cite{Lud02,Ens02}. In Fig. 2 we
plot the echo amplitude measured for the borosilicate glass BK7 as a
function of the magnetic field. At small $B$ the data show a few
oscillations; at higher fields the amplitude increases strongly and would
seem to saturate. The solid line for $B<100$ mT is calculated from Eq. (\ref
{eq30}) that, with $t_{\mathrm{w}}=2\mu $sec, $g=1.8$ and $I=\frac{3}{2}$,
describes the first minimum unambiguously. The increase at $B>100$ mT has
been fitted with (\ref{eq40}).

(ii) The particular waiting time dependence of the echo amplitude through
the product $Bt_{\mathrm{w}}$ has been verified experimentally in great
detail, especially for KBr:CN \cite{Ens02}. In Table I we give the nuclear
spin parameters $I$ and $g$ and the measured and calculated positions of the
first minimum in terms of the quantity $B_{\min }t_{\mathrm{w}}$.
Theoretical values are given for $p(x)=3x^{2}$ as discussed above \ (\ref
{eq34}). (The calculated values depend weakly on $p(x)$ and $b_{\beta }$.)

(iii) All systems showing the magnetic-field dependence\ contain nuclei with
$I\geq 1$ and finite quadrupole moments (B in BK7; Br and N in KBr:CN; Al in
Ba-Al-silicate). On the other hand, the only glass that shows no
magnetic-field dependence (amorphous silicon oxide) \cite{Lud-unp}, does not
contain nuclear quadrupoles, since $I=\frac{1}{2}$ for $^{29}$Si and $I=0$
for $^{16} $O and $^{28}$Si.

\begin{table}[h]
\caption{Nuclear spin parameters $g$ and $I$. Experimental values for the
first minimum of the echo amplitude. Calculated values as explained in the
text.}
\label{table1}\renewcommand{\arraystretch}{1.2} \setlength\tabcolsep{4pt}
\begin{tabular}{@{}lrclcc}
\hline
\noalign{\smallskip} &  &  &  & (meas.) & (calc.) \\
&  & $I$ & {\ }\ g{\ } & $B_{\mathrm{min}}t_{\mathrm{w}}\!\!$ & $\!\!$(10$%
^{-9}\,$Ts) \\ \hline
\noalign{\smallskip} borosilicate & $^{11}$B{\ }{\ }{\ } & 3/2{\ }{\ } & {\ }%
1.8{\ } & 22 & 21 \\
Al-Ba-silicate & $^{27}$Al{\ }{\ }{\ } & 5/2{\ }{\ } & {\ }1.44{\ } & 28 & 22
\\
KBr:CN & $^{79,81}$Br{\ }{\ }{\ } & 3/2{\ }{\ } & {\ }1.5{\ } & 30 & 28 \\
\hline
\end{tabular}
\end{table}

These findings provide very strong evidence that the observed magnetic-field
dependence arises from the dynamic phase of the nuclear Zeeman splitting of
quadrupolar levels. (Preliminary experiments on other systems would seem to
confirm this statement \cite{Lud-unp}.) The simplifications of the present
theory may be at the origin of the discrepancies in Fig. 2. For example,
real tunnel systems certainly involve more than one nuclear spin. This is
obvious for glasses.\ In mixed crystals, the tunneling atom drags its
dressing cloud; that's why bromine appears in Table I.

It seems likely at this point that the quadrupole splitting of tunneling
levels not only influences the amplitude of polarization echoes, but has
also consequences for other properties of glasses at very low temperatures.
In particular, we expect that the magnetic field dependence of the
dielectric susceptibility observed in several glasses is also caused by
nuclear spins.

In summary, we have proposed an explanation for the recently observed
magnetic-field dependence of polarization echoes in terms of a novel
dephasing mechanism involving nuclear spins and quadrupole moments. The EFG $%
\phi ^{\prime \prime }$ and the corresponding axes\textbf{\ }$\mathbf{u}$
are not the same in the two minima of the atomic double-well potential; the
quadrupolar energies gives rise to phase dispersion that reduces the echo
signal. The oscillations at small fields result from the interference of the
dynamical quantum phases of almost degenerate quadrupole levels. The strong
increase of the echo amplitude at larger $B$ is due to the alignment of the
nuclear magnetic moments with respect to the magnetic field. For strong
fields, $\mu _{\mathrm{N}}B\gg \hbar \omega _{\mathrm{Q}}$, we expect
saturation at the value $f=1$.

\end{document}